\documentclass[12pt]{article} 
\input{epsf.tex}
\usepackage{epsfig}
\newcommand{\One}{1\kern-4.5pt1}

\begin{document}
\vspace{-20pt}
 
\addtolength{\baselineskip}{0.20\baselineskip}

\newcommand{\re}{\mathop{\rm Re}}
\newcommand{\im}{\mathop{\rm Im}}
\newcommand{\tr}{{\rm Tr\,}}
\newcommand{\ex}{{\rm e}}
\newcommand{\bfx}{{\bf x}}
\newcommand{\htm}{\hat{T}}
\newcommand{\nn}{\nonumber}
\newcommand{\be}{\begin{equation}}
\newcommand{\ee}{\end{equation}}
\newcommand{\ba}{\begin{eqnarray}}
\newcommand{\ea}{\end{eqnarray}}
\newcommand{\eq}{Eq.~}
\newcommand{\eqs}{Eqs.~}
\newcommand{\fig}{Fig.~}

\begin{titlepage}
\begin{flushright}
CERN-PH-TH/2005-194 \\
CPT-2005/P-049\\
MS-TP-05-17
\end{flushright}
\begin{centering}
\vfill
                                                                                
{\bf \Large Screening of $Z(N)$-Monopole Pairs \\ in  Gauge Theories}
                                                                                
\vspace{0.8cm}
                                                                                
Ph. de Forcrand$^{1,2}$, C. P. Korthals Altes$^3$ 
and O. Philipsen$^4$
                                                                                
\vspace{0.3cm}

{\em $^{\rm 1}$
Institut f\"ur Theoretische Physik,
ETH Z\"urich,
CH-8093 Z\"urich,
Switzerland\\}
{\em $^{\rm 2}$
Theory Division, CERN, CH-1211 Geneva 23,
Switzerland\\}
{\em $^{\rm 3}$
Centre Physique Th\'eorique au CNRS,
Case 907, Campus de Luminy,\\ F13288 Marseille Cedex 9, France\\
}
{\em $^{\rm 4}$
Institut f\"ur Theoretische Physik, Westf\"alische Wilhelms-Universit\"at M\"unster, Germany}
                                                                                
\begin{abstract}\noindent
The screening of magnetic $Z(N)$-monopoles and the associated screening length in SU(N) gauge 
theories are analyzed theoretically, and 
computed numerically in the 3d $SU(2)$ theory. 
The nature of the screening excitations as well as their mass have so far
remained inconclusive in the literature.
Here we show that the screening mass is 
identical to the lowest $J^{PC}_R=0^{++}_+$ excitation of the 4d Yang-Mills Hamiltonian 
with one compact direction with period $1/T$, the subscript $R$ referring to
parity in this direction.
We extend the continuum formulation to the lattice, and determine the 
transfer matrix governing the decay of the spatial monopole correlator  
at any finite lattice spacing. 
Our numerical results for $SU(2)$ for the screening mass in the  dimensionally
reduced (high temperature) theory  are compatible with the  
$0^{++}$ glueball mass in 3d $SU(2)$. 

\end{abstract}
                                                                                
\end{centering}
\noindent
\vfill
\end{titlepage}

\section{Introduction}\label{sec:intro}

Ever since the recognition by 't Hooft and Mandelstam~\cite{tH76} that 
 confinement in QCD might be  a dual version of superconductivity, people
 have sought for  a quantitative understanding. The first step in that direction
 was the proof that electric confinement implied  the screening of colour magnetic fields~\cite{tH78}. That is of course a necessary condition:
the would-be magnetic Cooper pairs are expected to screen colour magnetic fields.
The proof in ref.~\cite{tH78} also pioneers the method that we will use below to extract the magnetic screening mass for any temperature, namely by introducing heavy magnetic sources and measuring their correlation. 't Hooft
used periodic boundary conditions and introduced sourceless fluxes winding around the periodic space directions. Electric confinement in this terminology means the electric fluxes carry an energy $F_e\sim L\sigma$. Here  $\sigma$ is the string tension per unit length and $L$ the size of the box in the direction of the flux. Now one makes the mild assumption that at zero temperature the energy of any array of electric and magnetic fluxes is additive, i.e. splits into a part $F_e$  due to electric fluxes and in a part $F_m$ due to magnetic fluxes. It then follows that $F_m\sim\sigma  L\exp(-\sigma L_{tr}^2)$, i.e. decays exponentially with the cross section orthogonal to the flux, the proportionality factor being the energy 
$\sqrt{\sigma}$ per unit length.  

Instead of studying periodic flux loops,
one may gain more information by looking at a monopole-antimonopole pair along the $z$-direction and by varying its separation $r$, in general at some finite temperature $T$. This yields a free energy $F_m(r,T)$ as a function of monopole separation, which is typically parametrized by a Yukawa behaviour, 
\begin{equation}
F_m(r,T)=F_0(T)-{c(T)\over r}\exp(-m_M(T)r).
\label{mass}
\end{equation}
Following the lattice formulation of the problem \cite{mack, dual}, there have been many attempts to compute the screening mass
$m_M$ by numerical simulations \cite{bil}-\cite{chernodub}, and to interpret the corresponding excitations  
as effectively massive 'constituent' gluons, Debye-screened gluons or glueballs.
Since an accurate numerical computation of the screening mass $m_M$ is 
very difficult, the numerical values are rather scattered and these studies have remained inconclusive.

In this paper we show that the
screening masses $m_M$ correspond to spin zero mass levels with $PC=1$  of a spatial Hamiltonian $H_T$.  This Hamiltonian propagates gauge invariant 3d Yang-Mills states (i.e.~glueballs) along the $z$-direction. The subscript $T$ means this Hamiltonian describes  
3d Yang-Mills fields in a box with the z direction periodic with period ${1\over T}$ and the $x$ and $y$ directions infinite in extent. So the rotation group in the $x-y$-plane is $SO(2)$. Let  parity $P$ be reflection in the two-dimensional infinite space, and $R$ be the reflection in the periodic direction. Together with charge conjugation $C$ the corresponding quantum numbers are conserved by $H_T$.
Until now a quantum number assignment  for the screening states was not mentioned in the literature, 
as far as we know~\footnote{Some of it is discussed by one of us in  the 2003 Zakopane lecture notes \cite{altespolo}.}. 
Here, we find that the assignments for the magnetic screening mass states $J^{P,C}_R$ are $J=0$ and $R=CP=1$.

Our underlying reasoning is similar to that applied to 
electric screening by Arnold and Yaffe some
time ago~\cite{arnold}. The operator  creating the heavy monopole is a local operator, 
in the same sense as the operator creating a heavy quark. 
The locality of the operator is somewhat masked in  the path integral representation, which 
may explain why it remained unnoticed. For this reason we will be 
fairly explicit. We discuss the lattice version of the operator in detail and establish that the relation between the 
screening mass $m_M$ and the mass of the lowest $0^{++}_+$  state 
is true for any finite lattice size $T^{-1}$.  
Two limiting cases are worth mentioning. At $T=0$ the screening mass $m_M$ corresponds to a 4d scalar glueball with $R=P$ because of full 3d rotation symmetry.
Since $CP=R=1$, we thus get the $0^{++}$ glueball mass. 
At large temperatures, modes of order $\sim T$ can be integrated out and screening masses are well described by the dimensionally reduced theory
(see \cite{dr} and references therein). 
The latter corresponds to a Yang-Mills theory for the colour magnetic fields $A_i\sim g^2T$ coupled to an adjoint scalar $A_0\sim gT$ in 3d. 
At $T>>m_D$, the Debye screening mass, the electric fields $A_0\sim gT$ may be integrated out as well, and our screening mass corresponds to a $0^{PC=1}$ state of the  2d Yang-Mills Hamiltonian, i.e. a glueball of the 3d Yang-Mills theory.

This connection is of great numerical value, since simulating the monopole pair is numerically quite involved\cite{rebbi,defor1}. Exploiting the connection, one might instead use a simple local source with the same quantum numbers as the heavy monopole and measure its correlation 
by well-established~\cite{oweowe}, numerical methods in a Yang Mills system 
at zero temperature and with one periodic direction. 

The layout of the paper is as follows. In section \ref{sec:creating}  monopole sources are discussed. Their correlation and the connection to the spatial
Hamiltonian is the subject of section \ref{sec:correlator} and section
\ref{sec:fictitious}.  In section \ref{sec:latticemethod} we will discuss the
lattice operator that excites the state with the screening mass, and the
relation of its correlator with the object we simulate: the twisted action. We identify the
transfer matrix whose lowest scalar glueball mass coincides with the screening
mass from our lattice correlator. Finally, in section \ref{sec:results}
 we will put this connection to the test by simulating at very high temperature the thermally reduced version of $SU(2)$ gauge theory. The restriction at very high $T$ permits us to simplify
the monopole correlator to one Euclidean time slice, i.e. to a two-point funtion.
Our results compare very well with the known masses~\cite{teper3d} in the corresponding 2d  reduced Hamiltonian.
Readers who are 
interested in the numerical results only may want to skip the first sections.

Before delving into the technical part of the paper a  word on motivation. The reader might wonder 
whether the study of magnetic screening is an entirely academic exercise.  We do
not believe so. Understanding  magnetic screening quantitatively  is a
 necessary step in the understanding of the magnetic activity that causes it. And
 this magnetic activity is at the core of confinement \cite{tH78}.  

Apart from this matter of principle there is a computational reason already partly mentioned before.  By rotating the time direction into a space direction one gets a spacelike 't Hooft loop, or ``electric flux loop'', with an area law behaviour. The loop tension
does not correspond to a level in the fictitious Hamiltonian $H_T$, although the simulation is identical in complexity~\cite{rebbi,defor1}. Thus a calibration of these simulation methods is more than  welcome and that is precisely provided by the connection of the magnetic mass with a mass level accessible through usual methods .

\section{Creating a heavy monopole}\label{sec:creating}   

In this section the analysis of the  creation operator of a Dirac monopole is resurrected. It may well have been done way back in the seventies, but we have been unable to find a reference. To a large extent it is related to the vortex 
operator in the seminal work of 't Hooft in 1978~\cite{tH78}, and used extensively in the work by Kovner \cite{kovner}.
We apologize for the pomp and circumstance that will go with this resurrection,
but we believe it is useful  to insist on the basic physics involved in order to
digest the sections that follow.

We consider an $SU(N)$ gauge theory in four dimensions, and matter fields 
that have $N$-ality zero. So they are neutral under the centergroup
$Z(N)$ of $SU(N)$. 
Consider  in such a theory the correlator of two heavy ``electric'' colour charges $Q_k$ with non-zero N-ality $k$. $Q_k$ is the creation operator of a particle field in a representation of SU(N) with N-ality $k$ and with an infinite mass. A centergroup transformation $\exp(i{2\pi\over N})$ transforms the source as follows :
\begin{equation}
Q_k\rightarrow \exp(ik{2\pi\over N}) Q_k.
\end{equation}
The correlator $\tr Q_k^{\dagger}(0,0,r)\exp(-H/T)Q_k(\vec 0)$ in path integral language  corresponds to the well known correlator of Polyakov loops,
\be
P(A_{\tau}(\vec x))={\cal{P}}\exp\left(i\int_0^{1/T}A_{\tau}(\vec x,\tau)d\tau\right),
\ee
up to obvious normalization factors:
\begin{equation}
\tr Q_k^{\dagger}(0,0,r)\exp(-H/T)Q_k(\vec 0)=\int DA \,P^{\dagger}(A_{\tau}(0,0,r))P(A_{\tau}(\vec 0)) \exp\left(-{1\over{g^2}}S(A)\right).
\label{esourcecor}
\end{equation} 
$A_{\tau}$ is in the same representation as $Q_k$.
The exponential decay of such a correlator, the static potential, 
is used as an order parameter:
at low temperature the potential rises linearly,
at high temperature it is screened like in eq.~(\ref{mass}).

Here we want to formulate its magnetic analogue. In this section we first
discuss the heavy monopole source, before using it to  
construct the magnetic correlator in the next section.

\subsection{Dirac monopole in electrodynamics}\label{subsec:dirac}

The monopole source in $SU(N)$ will be like the Dirac monopole in electrodynamics, which we will discuss first.
The simplest picture of a magnetic charge at the origin is one with a  radial magnetic field 
\begin{equation}
\vec B_{mon}=m{\vec x\over{4\pi r^3}}
\label{monopole}
\end{equation}
\noindent yielding a total charge $m$ from our monopole after integrating over any surface around the monopole. It is a solution of the Maxwell equation
$\vec\nabla.\vec B_{mon}=m\delta(\vec x)$.

Unfortunately the source term on the r.h.s. excludes the use of vector potentials because of the Bianchi identity, which says that a magnetic field expressed in terms of a gauge potential has no divergence. Dirac solved that problem by putting the source term at infinity. He then put a string of magnetic dipoles from that source to $\vec x=0$,
 guiding the flux through this thin string from the monopole at infinity to $\vec x=0$.
Let this  magnetic ``return flux'' come in along the $z$-axis $\vec\iota_z$. Then it is given by:
\begin{equation}
\vec B_s\equiv\nabla\times\vec S=-m\delta(x)\delta(y)\theta(z)\vec\iota_z.
\label{B_s_eq}
\end{equation}
Adding to  the monopole field strength  the return flux field strength,
\begin{equation}
\vec B_{mon}+ \vec B_s\equiv \vec B_{sol},
\end{equation}
gives a configuration with  divergence zero everywhere.                                          
 It is the field strength of a solenoid along the positive axis.
 By taking the surface integral
 it becomes explicit that the solenoid has no magnetic charge, because the integrated flux of the monopole field is neutralized by incoming flux of the string.  So there must be a potential $\vec A_{sol}$ with
 $\vec\nabla\times\vec A_{sol}=\vec B_{sol}$. Its precise form is not relevant for what follows.

Hence the
string represents the source of magnetic charge just as the delta function represents the source of electric charge. The end point of the string is where  the magnetic charge resides. If we are interested in having two monopoles of opposite sign then the string does not come in from infinity, but from the location of the second monopole with the opposite charge, and runs, as
before , to the monopole  eq. (\ref{monopole}) at the origin.

Of course the string should be invisible by scattering with a quantum mechanical particle with charge e. Then the string with its vector potential written as $\vec A_s$ causes a phase difference
\be
\exp(ie\oint d\vec l.\vec A_s).
\label{diraccond}
\ee
Now Stokes' theorem $\oint d\vec l.\vec A_s=\int d\vec S.\vec B_s=m$ tells us, that for the phase difference to vanish, the Dirac condition~\cite{dirac}  $em=2\pi n$ must apply. 

Finally we want to write down a monopole creation operator in QED. As we just saw, this amounts to creating an invisible string, obeying the Dirac condition.
The string has magnetic field zero everywhere except on the string itself, running along the positive $z$-axis. Thus the vector potential $\vec A_s$ of the string is a pure gauge  $\nabla \omega$ everywhere, except
 on the positive $z$-axis. There the gauge transformation is singular in such
a way as to recover the Dirac condition eq.(\ref{diraccond}). For any closed 
circuit around the string the gauge transform $ \omega$ has a discontinuity $\Delta\omega$ with: 
\be
e\oint d\vec l.\vec A_s=\Delta\omega=k2\pi.
\ee
So taking $\omega=k\phi$, the azimuthal angle around the string, we see that going around the string once, from $\phi=0$ to $\phi=2\pi$, the gauge transformation is forced to go  $k$ times around the circle that constitutes the gauge group $U(1)$.  Such transformations have a non-trivial homotopy $\Pi_1$. They are 
written as $\omega_k(\vec x_0;\vec x)$, $\vec x_0$ being the point where the string starts and runs along the positive $z$-axis.

Thus, to create a heavy monopole source in QED we have to effect a singular gauge transformation, generated by the Gauss operator $G(\vec x)=\nabla.\vec E -e\rho$, with charge density  $e\rho$.  A physical state in the Hilbert space is by definition invariant under regular transforms, in particular with trivial homotopy. So the monopole operator $M_1(\vec x_0)$ with charge $k=1$ in units of 
${2\pi\over e}$ equals:
\be
M_1(\vec x_0)=\exp{\left(i\int d\vec x \; \omega_1(\vec x_0;\vec x)G(\vec x)\right)}.
\label{monpoloqedcrea}
\ee

\subsection{Monopole source in SU(N) gauge theory}

Let us now turn to the generalization of the Dirac monopole to the SU(N)
 case. We take gluodynamics, possibly with matter fields in the adjoint representation.
The strength of the SU(N) Dirac monopole is again given by the discontinuity of the gauge transform encircling the string, the discontinuity now given by a centergroup element.
Because we suppose absence of charged Z(N) fields, the gauge group is really 
$SU(N)/Z(N)$. Encircling the string gives no singularity in this group, but corresponds to a non-contractible  path, just like for $U(1)$. 

Having identified the subgroup $Z(N)$ we now proceed to explicitly construct the operator that creates the  $Z(N)$ monopole.
We need to describe the string emanating from the monopole at the point $\vec 0$, say in the positive $z$-direction.  For that we need a gauge transformation 
$\Omega(\vec x)$ that has a discontinuity in the centergroup, when going around the string.  
To this end we define a function in the Lie-algebra of $SU(N)$: $\omega(\vec x)=\phi(\vec x){1\over N}Y_k$. $\vec x$ is any point on the string and  $\phi(\vec x)$ is the azimuthal angle  when following a closed curve around $\vec x$ in the $x-y$ plane.  $Y_k$ is a special traceless diagonal matrix in the Lie-algebra, which is the ``k-hypercharge''\cite{giovannaaltes}, $Y_k=\mbox{diag} (k,k,...,k,k-N,k-N,...k-N)$, such that when exponentiated
it gives a Z(N) group element:
\begin{equation}
\exp\left(i{2\pi\over N}Y_k\right)=\exp\left(ik{2\pi\over N}\right)1_{N\times N}.
\end{equation}
This means the diagonal $SU(N)$ gauge transformation 
\begin{equation}
\Omega_k= \exp\left(i\omega(\vec x){1\over N}Y_k\right)
\label{center}
\end{equation} 
picks up a discontinuity $\exp(ik{2\pi\over N})$,  when encircling the string.

The field strength of the string for the SU(N) Dirac monopole then reads $\vec B_s Y_k$ in the fundamental representation, with $B_s$ given by eq.(\ref{B_s_eq}).
There is an important question left. Under  regular gauge transformations this field strength transforms  as 
$$\vec B_s Y_k\rightarrow \Omega_r\vec B_s Y_k\Omega_r^{\dagger}.$$
Since a regular gauge transformation does not change the discontinuity, one 
would expect the total flux $\Phi$ of the monopole not to change. That means 
that  
the only gauge invariant flux one can have is $\exp(imY_k)$ with $m$, the strength of the  unit magnetic flux through the string, equal to ${2\pi\over N}$.

This motivates the definition of the Z(N) Dirac monopole operator with charge $\exp(ik{2\pi\over N})$ as the operator representation of the gauge transformation $\Omega_k$ in eq.(\ref{center}):    
\begin{equation}
M_k(\vec 0)=\exp\left(i(\vec E.\vec D\omega){1\over N}Y_k\right).
\label{monopolecreator}
\end{equation}
The dot means integration over the space coordinates and a trace over
colour.

The string will not affect physical states, built from $Z(N)$ invariant matter.
Physical states will only contain Wilson loops in $Z(N)$ neutral representations, and these loops will not respond to the string piercing them.
Therefore the location of the string is immaterial under these circumstances, only its endpoint matters. This means the monopole source is a scalar under rotations. Under P and C the orientation of the string reverses. The gauge invariant magnetic flux changes therefore from $\exp(i{2\pi\over N})$ to $\exp(-i{2\pi\over N})$.    The rotations and  parity discussed here are those of the 3d-space in which we live.
In section \ref{sec:fictitious} we will discuss the parity P and R parity in the fictitious 3d space with one direction compactified, and construct an operator that creates the corresponding state. 

At this point it is clear that 
the correlation of  two of our static magnetic scalar sources is analogous to that of  two static electric sources as in eq. (\ref{esourcecor}): we are looking at the magnetic Coulomb force and its screening. Note that this screening mass is different from the pole mass of the magnetic gluon propagator considered in other work \cite{karsch}.

\section{Correlator of two Z(N) monopoles at finite T}\label{sec:correlator}

Now that we have found the local operator that creates the monopole,
we want to know the force law between two of them. To formulate
 the force law, we start with a monopole at $\vec 0$, an anti-monopole at $(0,0,r)$ and form their correlation. This is the operator $M_k(0,r)$ given by a gauge transformation having the now familiar $Z(N)$ singularity     on a string going from  $\vec 0$ to $(0,0,r)$. Running it in imaginary time gives:
\begin{equation}
C(r)=<0|M_k(0,r)\exp{(-H\tau)}M_k^{\dagger}(0,r)|0>\equiv \exp(-\tau F_m(r,\tau))
\label{vvampl}
\end{equation}
\noindent for the vacuum to vacuum amplitude to create and annihilate our monopole pair. $F(r,\tau)$ 
for $\tau\rightarrow\infty$ is the energy associated with the pair.

Every Wilson loop with N-ality $l$ encircling the string will pick up a phase $\exp(ikl{2\pi\over N})$. $l$ is the number of fundamental representations the loop is made from. If the Hamiltonian contains a regularized version of the magnetic field strength density, as on the lattice, then:
\begin{equation}
B_z^2\rightarrow W_{xy}+W_{xy}^{\dagger}.
\label{hybrid}
\end{equation}
 
\begin{figure}
\begin{center}
\epsfig{file=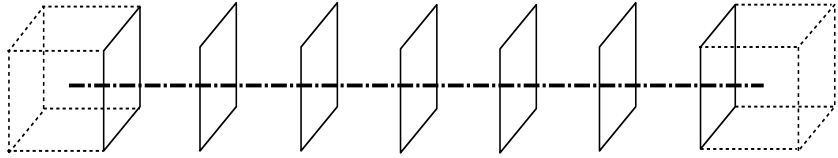,width=11cm,height=2cm}
\end{center}
\caption{\label{fig:chain}Monopole antimonopole pair induced by
twisting the (x,y) plaquettes pierced by the Dirac string in the $z$-direction.}
\end{figure}

This is a somewhat hybrid method, half continuum, half lattice, and we will give a full lattice version in section \ref{sec:latticemethod}.
The Wilson loop $W_{xy}$ is taken in the fundamental representation $l=1$. 
Then it is clear that for a fixed time slice  $\Delta\tau$ one can commute the correlator 
through the Hamiltonian factor $\exp(-H\Delta\tau)$ and produce a twisted Hamiltonian $H_k(r)$, with all loops encircling the string replaced by~\cite{groeneveld}:  
\begin{equation}
W_{xy}+W_{xy}^{\dagger}\rightarrow\exp{\left(ik{2\pi\over N}\right)} W_{xy}+\exp{\left(-ik{2\pi\over N}\right)}W_{xy}^{\dagger}
\label{twist}
\end{equation}
This is shown in fig. (\ref{fig:chain}).
Repeating this for all time slices, and going to the path integral representation of eq. (\ref{vvampl}) will give us the usual path integral in which the action $S$ is replaced by the twisted action $S_{(k)}(r,\tau)$\cite{groeneveld}:
\begin{equation}
\exp(-\tau F_m(r,\tau))\sim\int DA\exp\left(-{1\over{g^2}}S_{(k)}(r,\tau)\right)
\end{equation}
 The $xy$ plaquettes on
the sheet traced out by the string are all twisted like in eq.(\ref{twist}).
This creates a $(z,\tau)$ temporal 't Hooft loop.

The extension to finite temperature $T$ is now simply:
\begin{equation}
\exp\left(-{1\over T} F_m(r,\tau=1/T)\right)\sim\int DA\exp\left(-{1\over{g^2}}S_{(k)}(r,\tau=1/T)\right)
\label{ftcor}
\end{equation}
where the sheet now wraps around the full periodic time direction. In the following
we shall drop the argument $1/T$ in the action.
The free energy $F_m(r,\tau=1/T)$  depends on $T$ and is, as discussed above
eq. (\ref{mass}),  screened 
for {\it all} temperatures $T$.

\section{Space-periodic Hamiltonian and its symmetries}\label{sec:fictitious}

The correlation in its path integral form eq.(\ref{ftcor}) can be read as a propagator in the $z$-direction propagated by a Hamiltonian $H_T$. The ``time'' direction for this Hamiltonian is the $z$-direction. Its space directions are the $x, y$ and periodic $\tau$ direction. It correlates 
a ``magnetic Wilson line'' $V_k$ wrapping around the $\tau$-direction at $z=0$
with another one at $z=r$:
\ba 
C_{(k)}(r)&=&\int DA\exp(-{1\over{g^2}}S_{(k)}(r))\nn \\
  & \sim & \tr \exp{(-H_T (L-r))}V_k\exp{(-H_Tr)}V_k^{\dagger}
\label{ftcorz}
\ea
where $L\rightarrow\infty$ and the trace is over physical states.
The ``magnetic Wilson line'' $V_k$ is the operator
\begin{equation}
V_k=\exp\left(i\int^{1/T}_0 d\tau \vec{\tilde E}.\vec{ \tilde D}\omega(x,y){1\over N}Y_k\right),
\label{magneticline}
\end{equation}
 which is the analogue of the ``electric'' Wilson line $P(A_{\tau})$.
The tilde indicates that our canonical operators are now $A_{\tau,x,y}$ and $E_{\tau,x,y}$. The gauge function equals $\omega(x,y)=\arctan(y/x)$. Hence this operator creates for every value of $\tau$ a vortex  of strength $\exp(ik{2\pi\over N})$ in the origin of the $x,y$ plane. 

The Hamiltonian $H_T$ describes the Yang-Mills field in two infinite and one periodic dimensions and propagates it in the $z$-direction. For the  $z$-slice $\Delta z$ between $z=0$ and  $z=1$ the line operator $V_k$ will twist the $B_z$ operator in this Hamiltonian  over the full  period $1/T$ in $\tau$. This can be seen in the same fashion as in the previous section by commuting the $V_k$ operator through the
factor $\exp(-H_T\Delta z)$. Repeat this for the second slice and so on. We reproduce the same sheet of twisted $x-y$ plaquettes as in the previous section,  hence the identity eq.(\ref{ftcorz}). 

Inserting a complete set of states tells us that the lowest mass $m_L$ in 
the set of intermediate states will be the screening mass\footnote{Normalization is such that  the energy of the vacuum state is zero.}:
\begin{eqnarray}
<0|V_k\exp{(-H_Tr)}V_k^{\dagger}|0> & = & |<0|V_k|0>|^2+|<0|V_k|m_M>|^2{\exp(-m_Mr)\over r}+...\nonumber\\
                                    & = & \exp(-\frac{1}{T}F_0)\big(1+c_k{\exp{(-m_Mr)}\over r}+....\big).
\label{masshamilton}
\end{eqnarray}

If we want to create states with all momentum components zero, we have to
integrate the sources over $x$ and $y$ directions as well. This however would not
correspond to what we are doing on the lattice. Our source is fixed in $x$ and
$y$. Thus, in identifying from eq. (\ref{masshamilton}) the lowest mass, we have 
to integrate over momenta in the large directions $x$ and $y$. So that is why in that formula 
we have a Yukawa potential:
\be
|<0|V_k|m_M>|^2{\exp(-m_Mr)\over r},
\ee
resulting from the integral $\int {dp_xdp_y\over {E(p)}}\exp(-E(p)r)$.  
Neglecting momentum dependence in the matrix element is allowed at large enough $r$.
Now we have to identify the quantum numbers of $V_k$.    

\subsection{Quantum numbers}\label{subsec:quantumnumbers}

Consider the symmetry group ~\footnote{In fact this is not all of the symmetry group of $H_T$.
 There is the $Z(N)$ group generated by gauge transformations periodic modulo $Z(N)$ in the $\tau$ direction. They have as order parameter the Wilson line in the $\tau$ direction, $P(A_{\tau})$.  Below $T_c$ we have  $<0|P(A_{\tau})|0>=0$ whereas above $T_c$ $Z(N)$ is 
spontaneously broken: $<0|P(A_{\tau})|0>\neq 0$. Our operator $V_k$ does 
not transform under $Z(N)$, nor any local state. Only $P(A_{\tau})$ does.} of $H_T$:  
$SO(2)\times P\times C\times R$.
Apart from the 2d rotation group there is 2d parity $P$, which flips the sign of the $y$-axis, and $A_y\rightarrow -A_y$. There is charge conjugation $C$ with $A_{\tau,x,y}\rightarrow -A_{\tau,x,y}$, and $R$ parity changing $\tau\rightarrow {1\over T}-\tau$,
$A_{\tau}\rightarrow -A_{\tau}$. 

Clearly $V_k$ couples to scalars under SO(2), since it is  running parallel to the SO(2) rotation axis. Under $P$ and $C$ it transforms into $V_k^{\dagger}$, but $R$ parity leaves it invariant.  So our magnetic screening mass is a $0^{PC}_R$ state with $PC=R=1$. For the symmetric combination
$V_k+V_k^{\dagger}$ we have $P=C=1$, for the antisymmetric combination $P=C=-1$. 
Periodicity in the number of colours and charge conjugation tell us that $V_k$ and $V_{N-k}$ have the same effect.

However we can fix the assignment completely by reducing the temperature to 
$T=0$. Then we restore the full rotational symmetry. Since $R$ and $P$ reflections are related by the generator $J_x$ rotating the $y$-direction into the $\tau$ direction, a spin zero state like our magnetic mass state at $T=0$ must have $R=P$. That excludes the anti-symmetric combination mentioned before and only
$0^{++}$ is possible.

An amusing question comes up in connection with the correlator of $\im V_k$.
If it is to be different from  the correlator of ${\rm Re} V_k$, the correlator of $V_k$ should be non-zero. The question then arises how to implement the twisted
path integral version of such a correlator. Clearly one needs the return flux from a heavy monopole at infinity with strength $-2k$ (or $N-2k$). To simulate
such a twist configuration is in principle doable, but hard in practice~\footnote{We thank Christian Hoelbling for discussions on this point.}. But in practice one would -and could, as we just learnt- excite the same states by some local operator with the same quantum numbers as $\im V_k$.

It is instructive to compare with the electric Coulomb force, given by the Wilson line $P(A_{\tau}(\vec x))=\tr{\cal{P}}\exp(i\int_0^{1/T}d\tau A_{\tau}(\tau,\vec x))$.
$C$ and $R$ change it into $P(A_{\tau})^{\dagger}$, while $P$ leaves it invariant. So it excites $0^{PC}_R$ states with $CR=P=1$,
$C=R=1$ for the symmetric, $C=R=-1$ for the antisymmetric combination, and radial excitations. Like in the magnetic case we can narrow down the quantum numbers, as argued in ref.~\cite{arnold}. For small enough coupling we can use one loop perturbation theory, and the self-energy of $A_0$ is clearly odd in $R$.
Note a curiosity in the $SU(2)$ theory. There the imaginary part of $P(A_{\tau})$
 is zero, so it cannot excite states with $R=-1$. Nevertheless states with
$R=-1$ do exist in the spectrum of the periodic Hamiltonian $H_T$! Their mass
has been computed including  the first non-perturbative contribution\cite{hartowe,dr}. 
However, for all $SU(N)$ theories the correlator of $\im P(A_{\tau})$ is exponentially suppressed below $T_c$ (like $\exp(-{\sigma\over T} L_z)$, $\sigma$ being the string tension).

\subsection{Dimensionally reduced monopole correlator}\label{subsec:reducedcorrelator}

What happens under dimensional reduction to our monopole correlator? 
Obviously the one time slice version of our twisted action survives,
and all one has to do in a 3d simulation is to use the correlator along one 
string stretching from $\vec 0$ to $(0,0,r)$. 
The free energy $F(r)$ is then again a Yukawa potential (temperature dependence is absorbed into the parameters)
\begin{equation}
F_m(r)=F_0-{c\over r}\exp(-m_Mr).
\label{energy}
\end{equation}
The answer one gets in the continuum limit is expressed in terms of the 
3d coupling $g_M=g^2T$:
\begin{equation}
{m_M\over g^2_M}=d
\end{equation}
The dimensionless number $d$ 
gives the result for asymptotically high $T$. According to the identification
of the magnetic screening mass in the previous sections it should equal
the ratio of the $0^{++}_+$ mass over the dimensionful coupling $g_M^2$ in 3d Yang-Mills.
This is indeed what we find by numerical simulation and is discussed in
section \ref{sec:results}.

\section{Lattice formulation}\label{sec:latticemethod}

In this section we give the lattice formulation of the identification of the screening mass with the mass of a state of the periodic Hamiltonian. Indeed, as emphasised in section \ref{sec:correlator}, a consistent formulation requires a regulator for the magnetic field strength in the Hamiltonian, and a natural regulator is given by the lattice. In particular there is a subtlety with the choice of the transfer operator on the lattice. We will need a choice with the same spectrum as the conventional one, which at the same time produces the 
twisted lattice  action as defined in section  \ref{sec:correlator}.

For simplicity and numerical feasibility, we work in the dimensionally reduced formulation. At the end of this section the extension to the 3+1 dimensional case turns out to be quite straightforward.
We furthermore restrict our theoretical discussion to the simpler case where the spatial $(x-y)$-extent 
of the system is infinite. Our simulated volumes are chosen large enough so that finite volume effects
are negligible, as we will argue below.

\begin{figure}
\begin{center}
\epsfig{file=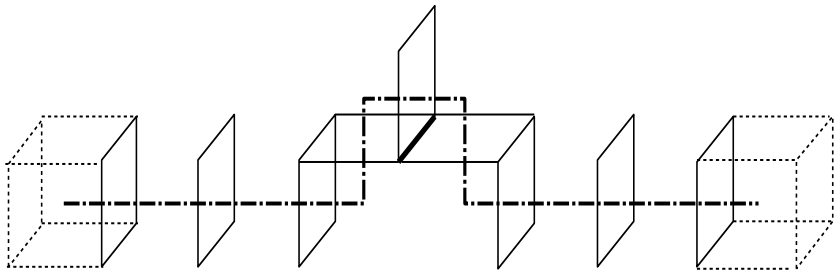,width=11cm,height=3.5cm}
\end{center}
\caption{\label{chain2}Change of the string location brought about by the
redefinition of the gauge variable on the link denoted by a bolder line.}
\label{fig:stringvar}
\end{figure}
Let M be the set of plaquettes being pierced by the Dirac string. They pick up a Z(N) factor $\zeta_k=\exp(ik{2\pi\over N})$. This is shown in Fig.~\ref{fig:chain}.
Then the lattice action for a Euclidean pure gauge theory in the presence of a
monopole pair separated by $r$ is
\be
S_{(k)}(r)=\frac{\beta}{N}\sum_{p\notin M}\re\tr U(p) +\frac{\beta}{N}\sum_{p\in M}\re \zeta_k \tr U(p),
\label{twisted_action}
\ee
which is obtained from the standard Wilson action $S$ by simply multiplying all
plaquettes $p\in M$ by the Z(N) factor $\zeta_k$. Correspondingly, this defines a partition function $Z_{(k)}(r)$ 
as a path integral evaluated with $S_{(k)}(r)$. We are interested in the behavior under
growing $r$, and thus consider the correlation function
\be
C_{(k)}(r)=\frac{Z_{(k)}(r)}{Z}=\left\langle \ex^{{\beta\over N}\sum_{p\in M}\re (\zeta_k - 1) \tr U(p)}\right\rangle \;.
\label{twistedcorrelator}
\ee

Before we carry on  we note that the location of the string in the path integral is not important. This is illustrated  in fig.~(\ref{fig:stringvar}). One takes a link variable $U(l)$ on any of the pierced plaquettes, and uses the invariance of the measure to change it into $\tilde U(l)=\zeta_k U(l)$. This has the effect shown in the figure: the original twisted plaquette loses the twist $\zeta_k$ when expressed in $\tilde U(l)$, but the other plaquettes bordering our link will acquire  the twist. A deformation in the string results, but the path integral remains the same. Only the end points of the string are fixed as the reader can easily verify.
This suggests that also on the lattice there is a {\it local} operator $V_L(\vec x_0)$ at the end points, creating and annihilating the vortex.  

For notational convenience we drop from now on the reference to the vortex center $\vec x_0$, and to the subscript $L$. We will write simply $V_k$ for  the operator $V_L$.
In continuum language this operator was a singular gauge transformation, 
and our task is now to find its lattice equivalent. More precisely, 
we want a lattice version of the Hamiltonian formula (\ref{ftcorz}), 
valid in the continuum:
\begin{equation} 
C_{(k)}(r)\sim  \tr \exp{(-H_T{L\over 2})}V_k\exp{(-H_Tr)}V_k^{\dagger}\exp{(-H_T({L\over 2}-r))}.
\end{equation}
In a lattice formulation we wish to express the r.h.s. in terms of some transfer matrix $T'$ and the  
 operator $ V_k$. In contrast to the standard formulation \cite{mack,dual}, 
we choose the transfer matrix to propagate states in the $z$-direction. 
That is, with the size $L_z$ in the $z$-direction 
and  distance $r$ in units of the lattice spacing, the correlator can be written as
\begin{equation} 
 C_{(k)}(r)\sim \tr T^{'{L_z\over 2}}V_k
 T^{'r}V_k^{\dagger} T^{'({L_z\over 2}-r)}.
\label{ftcorzlatt}
\end{equation}
In this form one has established exponential decay with the spectrum of $T'$ in the quantum number
sector of $V_k$, and the screening excitations are easily identified.

\subsection{The lattice vortex operator}

The vortex operator in  eq. (\ref{ftcorzlatt})  is given by its action on a state $|{U(l)}>$, 
where inside the ket the whole collection of $SU(N)$ link variables in a fixed time plane (with $x-y$ coordinates)  is given. The  operator valued  matrix $\underline U$  is diagonal in this basis:
\be
\underline U|U>=U|U>.
\ee
The action of the vortex operator on such a ket is simple. Draw from the vortex center in the middle of a plaquette, $\vec{x}_0=(x_0+a/2,y_0+a/2)$, a line to infinity, 
say parallel to the $x$-axis. This line will cross a set of links in the positive $y$-direction.
The vortex operator multiplies all those links with the Z(N) phase $\zeta_k=\exp(ik{2\pi\over N})$.
They are the "twisted" links. The rest of the links stays unaffected. We write in a shorthand notation:
\be
V_k^{\dagger}|U>=|\zeta_kU>.
\label{shorthand}
\ee
As we said explicitly before, the $x-y$ plane is taken to be infinite in extension. This avoids problems with the endpoint of the line that cuts the twisted links.
As a result, the plaquette containing the vortex at its center gets twisted, while all others remain untwisted.
Our  definition tells us that $V_k$ is unitary, just like the continuum operator. This can also be seen from the product $\langle U'|V_k^{\dagger}V_k|U\rangle=\langle zU'|zU\rangle=\delta(\zeta_k U',\zeta_k U)=\delta(U',U)$, 
since the delta function on the group is invariant under any rotation.  

An explicit representation of the vortex operator can be given in terms of the canonical variables
conjugate to the $\underline{U}$, the electric field operators defined by their commutation relations
$[{E}^a(l),E^b(l)]= if^{abc}E^c(l), [E^a(l),U(l)]= -T^aU(l) $ \cite{ham}. With $E^a(l)T^a=E(l)$, the vortex operator can then 
be written as \cite{mack}
\be
V_k=\Pi_l\exp\left(i\frac{4\pi}{N}\tr E(l)Y_k\right).
\ee
The product is over all links that are crossed by the line emanating from the vortex center $\vec{x}_0$. 

Now all Wilson loops in  the fixed $z$-plane that circumnavigate the vortex center will pick up the Z(N) phase. Hence, $V_k^{\dagger}$ transforms a physical state into another physical state. A physical state $|\psi>$ is a linear superposition of Wilson loop states $\sum_L\psi_L|L>$. 
Applying  $V_k^{\dagger}$ to such a state will change by a Z(N) phase those loops L that circumnavigate the vortex center, creating thereby a new physical state. 
It is then clear that the effect of $V_k^{\dagger}$ on a physical
state does {\it not} depend
on the way we defined the line of twisted links. This is another way of noting that a gauge transformation in one of the end points of a twisted link results in a deformation of the set of twisted links.

We thus expect the vortex operator to be a scalar under the remnant of the rotation group admitted by the lattice.
Under parity (in d=2: $y\rightarrow -y,x\rightarrow x$) and charge conjugation we have  $V_k\rightarrow V_k^{\dagger}$.
Hence in the continuum limit we expect two sets of excitations: a set $0^{++}$ and a set of $0^{--}$.
   
\subsection{The transfer matrix}

We define the transfer matrix $T=\exp(-aH_T)$ 
to propagate quantum mechanical states by one lattice spacing along the $z$-direction, i.e.~$z$
plays the role of time in our Hamiltonian treatment.
Starting from the conventional definition of $T$ \cite{ham},  we also introduce
a more unconventional $T'$ which is needed in order to write our correlator in the form eq.~(\ref{ftcorzlatt}). 
The difference between the two transfer matrices is in their decomposition  
into kinetic ($K$) and potential $(V)$ parts as
\ba
T&=&T_V^{1/2} T_KT_V^{1/2}\\
T'&=&T_K^{1/2}T_V T_K^{1/2}.
\label{transfer1}
\ea
Before defining the factors, we  
note that $T$ and $T'$ have the same spectrum, and hence they yield identical partition functions. 
This is inferred from the fact that the trace of any power of $T$ and $T'$ is the same, 
by using cyclicity of the trace:
\be
\tr T^N=\tr T'^N=\tr (T_VT_K)^N~~~~~~~~\mbox{for any N.}
\ee
The different factors are constructed from the potential and kinetic terms of the action, respectively. 
The potential factor $T_V$ is  a multiplication operator and 
consists of the exponent of the sum of all spatial ($x-y$) plaquettes $p_{ij}$:
\be
T_V=\exp\left({\beta\over N} \sum_{p_{ij}} \re \tr\underline U(p_{ij})\right).
\label{pot}
\ee
This operator is diagonal in the $U$-basis, $|U\rangle$.
The kinetic operator $T_K$ is an integral operator and defined by its matrix elements ($\vec r=(x,y),\vec\iota_i$ unit vector in $i$-direction, $i=x,y$):
\ba
<U'|T_K|U>&=&\int DU_z\exp\left(\frac{\beta}{N}\sum_{\vec r;i=x,y}\re \tr U_i(\vec r)U_z(\vec r+\vec\iota_i)U_i'^{\dag}(\vec r)U_z^{\dag}(\vec r)\right)\nn\\
&=&\int DU_z\exp\left(\frac{\beta}{N}\sum_{p_{0i}}\re \tr U(p_{0i})\right)
\label{tkdef}
\ea
On the r.h.s. of this definition appear all the "time"- like plaquettes $p_{0i}$ 
( i.e. with one link in the $z$-direction). In temporal (viz.~axial) gauge ($U_z=1$ on all $z$-links), 
this definition reduces to:
\be
<U'|T_{0K}|U>=\exp\left({\beta\over N}\sum_{\vec r;i=x,y}\re \tr U_i(\vec r)U_i'^{\dagger}(\vec r))\right).
\label{tkdef^}
\ee
The operators in axial gauge and the gauge invariant ones are connected
by a projection operator $P$,
\be
T_K=T_{0K}P=PT_{0K},  \qquad
 P = \int D g\,  R[g],
\ee
where the unitary operator $R[g]$ induces a gauge transformation with gauge function $g$,
$R[g]|U>=|U^g>$.

$T_V$ is a multiplication operator in the U-basis, so its square root is simply defined as:
\be
T_V^{1/2}=\exp\left({\beta\over{2N}} \sum_{p_{ij}} \re \tr\underline U(p_{ij})\right).
\label{sqrttv}
\ee
On the other hand, from our definition of $T_K$ and $T_{0K}$ it is not immediately obvious  
how to define their square root, since they are non-diagonal in the $U$-basis. 
A possibility is to define $T_{0K}^{1/2}$ by its matrix elements in terms of the 
character expansion.
The character $\chi_R(U)$ is the trace of the $d_R$-dimensional representation matrix
$D^R(U)$ in the given representation $R$, $\chi_R(U)=TrD^R(U)$.
The character expansion of the kinetic factor is
\be
\langle U'|T_{0K}|U\rangle=\sum_Rd_R\chi_R(U U'^{\dag})I_R(\beta).
\label{kinex}
\ee
Assuming that the square root is also a function of $\re \tr (U(l_i)U'^{\dagger}(l_i))$, 
its character expansion is
\be
\langle U'|(T_{0K})^{1/2}|U\rangle=\sum_Rd_R\chi_R(UU'^{\dag})J_R(\beta),
\label{kinexroot}
\ee
with as yet unknown coefficients $J_R$.
To find these we note that
the square root of the kinetic factor is required to satisfy
\be
( T_{0K})^{1/2}( T_{0K})^{1/2}= T_{0K}.
\ee
In terms of the matrix elements this means
\be
\int dU''\langle U'|(T_{0K})^{1/2}|U''\rangle \langle U''|(T_{0K})^{1/2}|U\rangle=\langle U'|T_{0K}|U\rangle .
\label{unit}
\ee
Using eq.~(\ref{kinexroot})
and the orthonormality of the characters~\footnote{This formula follows from the orthogonality of the representations:
\be
\int dU D^R(U)_{m,n} D^{R'}(U^{\dagger})_{n',m'}={1\over{d_R}}\delta_{R,R'}\delta_{m,m'}\delta_{n,n'}\nonumber\\.
\ee},
\be
\int dU''\chi_R(UU'')\chi_{R'}((U'U'')^{\dagger})=\delta_{R,R'}{1\over{d_R}}\chi_R(UU'^{\dagger}),
\label{characcomp}
\ee
one finds that eq.~(\ref{unit}) reduces to:
\be
\langle U'|T_{0K}|U\rangle=\sum_Rd_R\chi_R(UU'^\dag)J_R^2(\beta).
\ee
Comparing coefficients with eq.~(\ref{kinex}) one determines
\be
J_R(\beta)^2=I_R(\beta).
\label{sqrt}
\ee
Now we define
\be
T_{K}^{1/2}\equiv T_{0K}^{1/2}P,
\ee
and observing that
 $T_{0K}^{1/2}P=PT_{0K}^{1/2}$, one easily verifies that $T_{K}^{1/2}T_{K}^{1/2}=T_{K}$ as desired.

\subsection{The twisted transfer matrix}

Next, we consider the action of the vortex operator on the transfer matrix. Acting on the 
kinetic operator in axial gauge, its matrix elements are
\be
<U'|V_k T_{0K} V_k^{\dagger}|U>=<\zeta_k U'| T_{0K}|\zeta_kU>,
\ee
according to the definition of $V_k$.
Using eq.~(\ref{tkdef^})  we get
\be
<\zeta_k U'| T_{0K}|\zeta_kU>
   =\exp{\beta\over N}\{ \sum_{l_t}\re \tr (\zeta_kU'(l'_t))^{\dagger}\zeta_kU(l_t)+\sum_{l_{nt}}\re \tr U'(l'_{nt})^{\dagger}U(l_{nt})\}.
\label{deff}
\ee
Only in the twisted links ($l_t$) there is the centergroup factor, not in the untwisted links $l_{nt}$.
Of course {\it all} the centergroup elements drop out of the r.h.s. of eq.~(\ref{deff}), so $V_k T_{0K} V_k^{\dagger}=T_{0K}$ . This is true for {\it any} lattice action defined in terms of a sum of  traces of irreducible representations.
Furthermore, the vortex operator commutes with the projection operator $P$, since the latter does not contain any link variables. Hence we also have
\be
V_kT_KV_k^{\dagger}=PV_kT_{0K}V_k^{\dagger}
               =PT_{0K}=T_K.
\ee

On the other hand, the potential factor of the transfer matrix contains the plaquette which encircles the
vortex center, and this plaquette picks up a twist. We therefore obtain the twisted potential operator
\be
T_V^{(k)}\equiv V_kT_VV_k^{\dagger}= \exp\left({\beta\over N} \sum_{p_{ij}\notin M} \re \tr\underline U(p_{ij})\right)
\,\exp\left({\beta\over N}  \re \zeta_k \tr\underline U(p_{ij}(\vec{x}_0))\right).
\ee
With these results the twisted transfer matrices take the form
\ba
T^{(k)}&=&V_kTV_k^\dag=T_V^{(k)1/2} T_KT_V^{(k)1/2}\nn\\
T^{'(k)}&=&V_kT'V_k^\dag=T_K^{1/2}T_V^{(k)} T_K^{1/2}.
\label{twistedt}
\ea
We are now able to express our lattice correlator eq.~(\ref{twistedcorrelator}) in terms of these quantities. Let the point $\vec x_0$ in the x-y plane  denote the midpoint of the twisted plaquette $p_{xy}(\vec{x}_0)$ with the weight  $\zeta_k U(p_{xy}(\vec{x}_0))$. The transfer matrix connects two such plaquettes in the z-direction (see fig.~\ref{fig:chain})). Then the correlator is: 
\ba
C_{(k)}(r)&=&Z^{-1}\tr \left\{T^{L_z-r} \,\left(\ex^{(\frac{\beta}{N}\re(\zeta_k -1)\tr U(p_{xy}(\vec{x}_0)))}\,
T\right)^{r}\right\}\nn\\
&=&Z^{-1}\tr \left\{T^{L_z-r-1}\,T_V^{1/2}T_KT_V^{(k)1/2}\;\left(T^{(k)}\right)^{r-1}\;
T_V^{(k)1/2}T_KT_V^{1/2}\right\}.
\ea
In this form  we have an unwanted endpoint effect in
the correlator, since the transfer matrices at the locations of the monopoles have only their inner
potential factor twisted. 
The problem is remedied by using $T'$ instead, in terms of which one simply gets
\be
C_{(k)}(r)=Z^{-1}\tr \left\{T^{'L_z-r}\,\left(T^{'(k)}\right)^r\right\}.
\ee
Using eq.~(\ref{twistedt}) we finally obtain our desired form for the correlator,
\be
C_{(k)}(r)=Z^{-1}\tr \left\{T^{'L_z-r}\,V_k\, T^{'r}\, V_k^\dag\right\}.
\ee
Since $T$ and $T'$ have the same spectrum, we thus conclude that the monopole pair 
correlator decays with distance, governed by the physical spectrum of the transfer matrix.
For large $r$, the screening mass $m_M$ should thus represent the lightest state of the spectrum that $V_k$
couples to, and in Yang-Mills theory this would be the $0^{++}$ glueball.

\subsection{Generalization to 3+1 dimensions.}\label{subsec:3+1}

A similar discussion applies to the case where a small periodic space dimension of length $1/T$ is added to our large two dimensional $x-y$ space. The line of twisted plaquettes is then extended to a  surface of twisted plaquettes which is closed in this periodic
direction and defines the twisted action $S_{(k)}$ in eq.(\ref{ftcor}).

 In the operator formalism our vortex operator is then repeated for every slice along a closed loop in this 
periodic direction and the product of all is denoted by $V_k$. 
Closure of this surface in the periodic direction guarantees that no plaquettes with a link in the periodic direction will be twisted.
Choosing the transfer matrix $T'$ 
of eq.~(\ref{transfer1}) for the correlator of the extended vortex operators reproduces the path integral with the twisted action $S_{(k)}$.  Finally, the length $1/T$ can be taken to infinity.

In conclusion, in an infinite volume the correlator eq.(\ref{twistedcorrelator}) is measuring the lowest  mass of the states  excited by the vortex operator.
This is true for any finite lattice spacing.
In SU(2) there is only one set of states  excited, and the $0^+$ is the lowest one in the continuum limit.

\section{Lattice results in 3d}\label{sec:results}


The goal of our numerical study is to measure the screening mass $m_M$ from
the large-distance behaviour of the monopole correlation $C_k(r)$, eq.~(\ref{twistedcorrelator}).
We work in the infinite temperature limit, where the system is effectively 3d and the temperature
dependence of the free energy can be dropped, so
\be
C_{(k)}(r)=\frac{Z_{(k)}(r)}{Z} = e^{-F_m(r)/T}, \quad F_m(r) = F_0-{c\over r}\exp(-m_M r),
\ee
(cf eq. (1) and (18)) corresponding to the exchange of a boson of mass $m_M$ in 3 dimensions. 
$Z_{(k)}(r)$ is the 
partition function of a system with a stack of $k$ twisted (or ``flipped'' for $SU(2)$)
plaquettes as per eq.(\ref{twisted_action}).

To compute $\frac{Z_{(k)}(r)}{Z}$, our strategy consists of factorizing the ratio into $r$
factors, each of order 1, as done in Ref.\cite{defor1} for the 4-dimensional case:
\be
\frac{Z_{(k)}(r)}{Z} = \frac{Z(r)}{Z(r-1)} \times \frac{Z(r-1)}{Z(r-2)} \times .. \times \frac{Z(1)}{Z},
\ee
where $Z(r) = Z_{(k)}(r)$ and $Z(j)$ has $j$ twisted plaquettes. Each factor can be written as an expectation
value:
\be
\frac{Z(j)}{Z(j-1)} = \left\langle \exp( \frac{\beta}{N} \re \tr (\zeta_k - 1) U({p_{j}}) )\right \rangle ,
\label{obs1}
\ee
where $U(p_{j})$ is the $j^{th}$ plaquette to be twisted, and the expectation value
is taken with respect to $Z(j-1)$. Equivalently, the same ratio can be expressed as a ratio of
two expectation values with respect to an interpolating partition function $Z(j-1/2)$, where
$U({p_j})$ is multiplied by $\frac{\zeta_k + 1}{2}$:
\be
\frac{Z(j)}{Z(j-1)} = \frac{\langle \exp( \frac{\beta}{N} \re \tr \frac{\zeta_k - 1}{2} U({p_j}) ) \rangle_{j-1/2}}
                           {\langle \exp( \frac{\beta}{N} \re \tr \frac{1 - \zeta_k}{2} U({p_j}) ) \rangle_{j-1/2}}.
\label{MCobs}
\ee
This latter expression has a smaller variance than eq.(\ref{obs1}). Moreover, for $SU(2)$
$\frac{\zeta_k + 1}{2} = 0$, so that plaquette $p_j$ has zero coupling in $Z(j-1/2)$.
Since the 4 links making up $U({p_j})$ are thus decoupled, one can form $n^4$ estimates of the
two observables, from $n$ estimates of each of the 4 links. This provides additional variance reduction.

\begin{figure}[th]
\begin{center}
\epsfig{file=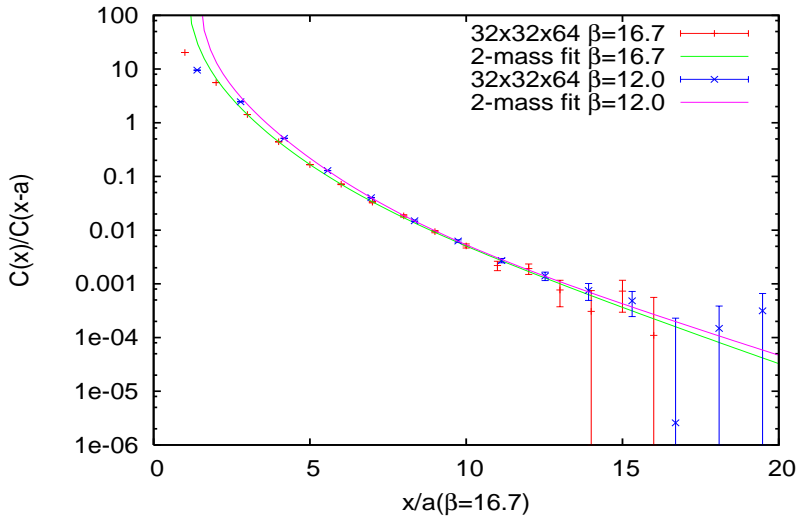,width=11cm,height=7.0cm}
\end{center}
\caption{Monte Carlo measurements of $C(x)/C(x-a)$ (eq.(\ref{MCobs})) versus distance.
Simulations at $\beta=16.7$ and $12.0$ give consistent results.
A 2-mass fit per eq.(\ref{fit_ansatz}) describes the data well, from $x/a=3$ on.}
\label{fig_data}
\end{figure}

On a lattice of size $32^2 \times 64$, for each value of $j=1, 2, .., 32$ (or in practice until the signal
became undetectably small), we have performed independent Monte Carlo simulations, collecting
$1-3~10^6$ measurements in each. 
Each simulation provides an estimate of 
$\frac{Z(j)}{Z(j-1)} = \frac{C(ja)}{C(ja-a)}$
for a different $j$.
We can then compare all these estimates to the functional form 
\be
\log \frac{C(r)}{C(r-a)} = \frac{\sum_{i=0}^{i_{\rm max}} c_i/r \exp(-m_i r)}{\sum_{i=0}^{i_{\rm max}} c_i/(r-a) \exp(-m_i (r-a))},
\label{fit_ansatz}
\ee
corresponding to the exchange of $(i_{\rm max} + 1)$ increasingly heavy scalar bosons of masses $m_i$.
The successive ratios $\frac{C(r)}{C(r-a)}$ are shown in Fig.~\ref{fig_data}.
A single boson exchange is insufficient to describe the data. 
Therefore, we fit the data to an Ansatz corresponding to the exchange
of 2 bosons of masses $m_0$ and $m_1$, over the range $r \in [3,20]$. 

\begin{table}[th]
\begin{center}
\begin{tabular}{|c|c|c|}
\hline
$\beta$ & $m_0/g_M^2$ & $m_1/g_M^2$ \\
\hline
16.7    & 1.73(12)  & 3.75(19)  \\
12.0    & 1.54(10)  & 3.43(10)  \\
\hline
together & 1.69(8)  & 3.67(12)  \\
\hline
\end{tabular}
\end{center}
\caption{Mass of the lightest exchanged boson ($m_0/g_M^2$) and next-lightest ($m_1/g_M^2$)
from simulations at two different values of the lattice spacing in the scaling region,
corresponding to the curves Fig.~\ref{fig_data}.
The two sets of results are consistent and can be fitted together (bottom line).}
\end{table}

The fitted masses are shown in Table I. To check that we are simulating continuum
physics, we repeated the simulations at two different values of the lattice spacing,
corresponding to $\beta=16.7$ and $12.0$. The fitted masses from each simulation are consistent.
Thus, we can fit both datasets together, obtaining the last line of Table I.
The groundstate mass $m_0/g_M^2 = 1.69(8)$ is compatible with the mass of the $0^{++}$
glueball measured in the zero-temperature $3d$ SU(2) theory~\cite{teper3d},
$m(0^{++})/g_M^2 = 1.58(2)$. We interpret the fact that it is somewhat larger as being due to remaining contamination of higher excitations (for example, no zero momentum projection is used in  the monopole correlator.)
On lattices of the sizes used here, glueball masses are known to be free of finite size effects, so we believe our infinite volume discussion is justified.

\section{Summary and prospects}

In this paper we have determined some simple but useful properties of
the magnetic Z(N) screening mass.  In particular, we have shown that the screening mass has the numerical value of the $0^{++}_+$ state of a spatial Hamiltonian
with a periodic space dimension of length $1/T$. So the screening mass can be computed by correlating simple local operators with the right quantum numbers. For the extreme case of infinitely large T we have shown by lattice simulation that indeed the  magnetic mass as measured by twisted actions and
the $0^{++}_+$ state in 2+1 dimensions do coincide. 
The latter is very encouraging for the technique of simulating the twisted action in cases where one cannot revert to a simpler operator, as in ref.~\cite{defor1}.

It is very desirable to evaluate the magnetic mass for all  T between the two cases where they are already known.  How sensitive is the magnetic mass to the Z(N) transition that governs the behaviour of the thermal Polyakov line? One would say that the real part of the Polyakov line has the required quantum numbers to excite the $0^{++}_+$. Thus one might be led to think that the magnetic mass
should follow the behaviour of the Polyakov line: second order in SU(2), almost second order in SU(3), and first order for N larger than 3. 
However, Z(N) symmetry operators do {\it not} affect local states . Only below $T_c$
 there are torelon states (i.e. string states winding in the periodic
direction~\cite{tH78}), that do transform non-trivially, and indeed are disappearing above $T_c$. 
Ref.~\cite{defor1} shows that the screening mass smoothly increases with temperature, 
with no special feature at $T_c$. However, for SU(3) the mass of the $A_0^2$
representing the $0^{++}$ ground state dips somewhat as one approaches $T_c$ \cite{oweowe}.  
We thus feel the jury is still out on the behaviour of magnetic screening and the spatial string tension near the critical point.

\section{Acknowledgements}
CKA would like to thank Jan Smit for discussions on the lattice formulation, Mikko Laine for advice, and the Instituut Theoretische Fysica Amsterdam for its hospitality.
We are all indebted to Christian Hoelbling for collaboration in an early stage of this work, and
to the Kavli Institute of Theoretical Physics, UCSB, for hospitality, funding and a 
stimulating atmosphere.


\begin{thebibliography}{xx}
\bibitem{tH76} G.~'t Hooft, in High Energy Physics, Ed. A. Zichichi (Editrice
Compositori Bologna, 1976);
S.~Mandelstam, Phys.~Rep.~23C (1976), 245.
\bibitem{tH78} G. 't Hooft, Nucl.Phys.B138, 1, (1978); Nucl.Phys.B153:141,(1979) .
G.~'t Hooft,
  Nucl.\ Phys.\ B {\bf 138} (1978) 1;
  Nucl.\ Phys.\ B {\bf 153} (1979) 141.
\bibitem{mack}
G.~Mack and V.~B.~Petkova,
  Annals Phys.\  {\bf 123} (1979) 442;
  Z.\ Phys.\ C {\bf 12} (1982) 177.
\bibitem{dual}
 A.~Ukawa, P.~Windey and A.~H.~Guth,
  Phys.\ Rev.\ D {\bf 21} (1980) 1013.
\bibitem{bil}
 A.~Billoire, G.~Lazarides and Q.~Shafi,
  Phys.\ Lett.\ B {\bf 103} (1981) 450.
\bibitem{deg}  
  T.~A.~DeGrand and D.~Toussaint,
  Phys.\ Rev.\ D {\bf 25} (1982) 526.
\bibitem{tep}
A.~Hart, B.~Lucini, Z.~Schram and M.~Teper,
  JHEP {\bf 0011} (2000) 043
  [arXiv:hep-lat/0010010].
\bibitem{rebbi}
C.~Hoelbling, C.~Rebbi and V.~A.~Rubakov,
  Phys.\ Rev.\ D {\bf 63} (2001) 034506
  [arXiv:hep-lat/0003010].
\bibitem{defor1}
  P.~de Forcrand, M.~D'Elia and M.~Pepe,
  Phys.\ Rev.\ Lett.\  {\bf 86} (2001) 1438
  [arXiv:hep-lat/0007034].
\bibitem{deForcrand:2001nd}
  P.~de Forcrand and L.~von Smekal,
  Phys.\ Rev.\ D {\bf 66} (2002) 011504
  [arXiv:hep-lat/0107018].
\bibitem{chernodub}
  M.~N.~Chernodub, F.~V.~Gubarev, M.~I.~Polikarpov and V.~I.~Zakharov,
  Phys.\ Lett.\ B {\bf 514} (2001) 88
  [arXiv:hep-ph/0101012].
\bibitem{arnold}
P.~Arnold and L.~G.~Yaffe,
  Phys.\ Rev.\ D {\bf 52} (1995) 7208
  [arXiv:hep-ph/9508280].
 \bibitem{dr}
A.~Hart, M.~Laine and O.~Philipsen,
  Nucl.\ Phys.\ B {\bf 586} (2000) 443
  [arXiv:hep-ph/0004060].
 A.~Hart and O.~Philipsen,
  Nucl.\ Phys.\ B {\bf 572} (2000) 243
  [arXiv:hep-lat/9908041].
\bibitem{altespolo}
  C.~P.~Korthals Altes, Zakopane Lectures,
  arXiv:hep-ph/0308229.
\bibitem{oweowe}
For a review, see:
E.~Laermann and O.~Philipsen,
  Ann.\ Rev.\ Nucl.\ Part.\ Sci.\  {\bf 53} (2003) 163
  [arXiv:hep-ph/0303042].
M. Laine, SEWM 2002  Proceedings, Ed. M.Schmidt, World Scientific,
  arXiv:hep-ph/0301011.
  \bibitem{teper3d} 
  M.~J.~Teper,
  Phys.\ Rev.\ D {\bf 59} (1999) 014512
  [arXiv:hep-lat/9804008].
  \bibitem{kovner}
  A.~Kovner,
  Int.\ J.\ Mod.\ Phys.\ A {\bf 17} (2002) 2113
  [arXiv:hep-th/0211248].
\bibitem{dirac} 
P.~A.~M.~Dirac,
  Proc.\ Roy.\ Soc.\ Lond.\ A {\bf 133} (1931) 60.
\bibitem{giovannaaltes}
P.~Giovannangeli and C.~P.~Korthals Altes,
  Nucl.\ Phys.\ B {\bf 608} (2001) 203
  [arXiv:hep-ph/0102022].
  \bibitem{karsch}   
  W.~Buchmuller and O.~Philipsen,
  Nucl.\ Phys.\ B {\bf 443} (1995) 47
  [arXiv:hep-ph/9411334].
  Phys.\ Lett.\ B {\bf 397} (1997) 112
  [arXiv:hep-ph/9612286].
 A.~Cucchieri, F.~Karsch and P.~Petreczky,
  Phys.\ Rev.\ D {\bf 64} (2001) 036001
  [arXiv:hep-lat/0103009].
O.~Philipsen,
  Phys.\ Lett.\ B {\bf 521} (2001) 273
  [arXiv:hep-lat/0106006];
  Nucl.\ Phys.\ B {\bf 628} (2002) 167
  [arXiv:hep-lat/0112047].
  \bibitem{groeneveld}
   J.~Groeneveld, J.~Jurkiewicz and C.~P.~Korthals Altes,
  Phys.\ Lett.\ B {\bf 92} (1980) 312.
\bibitem{hartowe}
K.~Kajantie, M.~Laine, J.~Peisa, A.~Rajantie, K.~Rummukainen and M.~E.~Shaposhnikov,
  Phys.\ Rev.\ Lett.\  {\bf 79} (1997) 3130
  [arXiv:hep-ph/9708207].
  M.~Laine and O.~Philipsen,
  Phys.\ Lett.\ B {\bf 459} (1999) 259
  [arXiv:hep-lat/9905004].
\bibitem{ham}
J.~B.~Kogut and L.~Susskind,
Phys.\ Rev.\ D {\bf 11} (1975) 395.
M.~Luscher,
Commun.\ Math.\ Phys.\  {\bf 54} (1977) 283.
M.~Creutz,
Phys.\ Rev.\ D {\bf 15} (1977) 1128.
J.Smit, Introduction to Field Theory on a Lattice, Cambridge University Press;
M.Creutz, Quarks and Gluons on the Lattice, Cambridge University Press.

  
\end{thebibliography}
\end{document}